%% file: ms.tex
\documentstyle[aaspp4]{article}
 
\lefthead{Schneider \& Schombert}
\righthead{Nearby LSB Galaxies}
 
\begin{document}
 
\title{Nearby Gas-Rich Low Surface Brightness Galaxies} 

\author{Stephen E. Schneider}
\affil{Astronomy Program, University of Massachusetts,
    Amherst, MA 01003}

\author{James M. Schombert}
\affil{Department of Physics, University of Oregon}
 
\begin{abstract}

We examine the Fisher--Tully $cz<1000$ km s$^{-1}$ galaxy sample to determine
whether it is a complete and representative sample of {\it all} galaxy types,
including low surface brightness populations, as has been recently claimed. 
We find that the sample is progressively more incomplete for galaxies with (1)
smaller physical diameters at a fixed isophote and (2) lower HI masses. This
is likely to lead to a significant undercounting of nearby gas-rich low
surface brightness galaxies. However, through comparisons to other samples we
can understand how the nearby galaxy counts need to be corrected, and we see
some indications of environmental effects that probably result from the local
high density of galaxies.

\end{abstract}
 
 
\section{Introduction}

Two of the key questions in 20th century extragalactic studies have concerned
the density and the composition of the galaxy population in the Universe. Our
knowledge of galaxy types and their abundance depends critically on the issue
of completeness of our galaxy catalogs. Since our current catalogs are
constructed by various observational means, they are, by definition, limited by
natural and technological selection effects.

Recently, Briggs (1997a) has argued that the Fisher \& Tully (1981, F--T)
catalog of nearby galaxies is complete to its redshift and sensitivity limits,
even for low surface brightness (LSB) galaxies. F--T examined HI emission from
a sample of 1787 angularly large galaxies accessible from the Green Bank radio
telescopes. They believed the sample to be very complete for late-type galaxies
within a redshift of $cz=1000$ km s$^{-1}$, with angular diameters larger than
3$'$, located in regions at $|b|>30^\circ$ and $\delta>-33^\circ$. Briggs
additionally found a sensitivity limit depending on the HI mass and defined the
F--T ``completeness zone'' as extending out to:
\begin{equation}
z_{CZ} = \left\{ \begin{array}{ll} 1000\mbox{km s}^{-1}/c & 
\mbox{if $M_{HI}>10^{8.45} h_{75}^{-2} M_\odot$} \\
(M_{HI}h_{75}^{2}/10^{8.45}M_\odot)^{5/12} 1000\mbox{km s}^{-1}/c &
\mbox{otherwise.}
\end{array} \right.  
\end{equation} 
The form of the limit for sources with masses smaller than 
$10^{8.45}M_\odot$\footnote{We adopt $H_0=75$ km s$^{-1}$ Mpc$^{-1}$ for masses
and distances quoted hereafter, and we use the heliocentric velocity corrected
by 300 km s$^{-1} \cos b \sin l$ as in F--T and Briggs.} was based on a
semi-analytic, semi-empirical fit to the HI sensitivity.

Briggs pointed out that surveys of low surface brightness galaxies (Schneider
et al.~1990; Schombert et al.~1992; Matthews \& Gallagher 1996; Impey et
al.~1996) have identified sources that are primarily at larger distances, but
they have added very few within the F--T ``completeness zone.'' He concluded
that LSB galaxies ``must already be fairly represented by nearby, previously
cataloged examples.''

This is an interesting idea, but the conclusion does not necessarily follow
from the analysis for several reasons: (1) For sources with low HI masses
$z_{CZ}$ is so small that redshift distances are very uncertain and Galactic
HI emission creates strong confusion. A further objection to Briggs' analysis
is that (2) the F--T sources are themselves incomplete within Briggs'
``completeness zone'' because of angular size selection effects.  Lastly, (3)
the $cz<1000$ km s$^{-1}$ region around the Local Group has about twice the
average galactic density and therefore is not a very representive sample of the
Universe as a whole.

In a companion paper, Briggs (1997b) used the F--T and LSB samples to derive
an HI mass function. He commented on the need for corrections for
incompleteness and noted the usefulness of a  ${\cal V}/{\cal V}_{max}$ test
for establishing whether the LSB samples were complete, but he did not discuss
the problems the F--T sample has in this regard. In this paper we explore the
limitations of the F--T sample and discuss how it and more recent surveys may
be properly used to understand the composition of galaxy populations. We show
that the local samples of galaxies display morphological segregation
characteristics associated with high density environments. Finally, we find
that in all HI mass ranges the dominant class of galaxies are those with the
smallest angular diameters at the isophotal limits of the original Palomar Sky
Survey (PSS-I). These galaxies may be physically small or appear small because
they are LSB systems; in either case, they are greatly under-represented in
the F--T sample and most other optical surveys.

\section{Completeness Tests of the F--T Sample}

The incompleteness of the F--T sample can be demonstrated using a ${\cal V} /
{\cal V}_{max}$ test.  This test compares the distance $d$ of a detected source
to the maximum distance $d_{max}$ at which it should be detectable. If the
maximum distance is correctly estimated, a source is equally likely to fall
anywhere within the volume delimited by $d_{max}$. On average, then, sources
will be found halfway into the maximum volume, and ${\cal V} / {\cal V}_{max}
\equiv (d/d_{max})^3$ will average 0.5 (Schmidt 1968). For a sample of $N$
sources, the probability distribution of the mean value of ${\cal V}/{\cal
V}_{max}$ has a nearly normal distribution with standard deviation
$1/\sqrt{12N}$.

We assume the distance is proportional to the redshift $z$, so that ${\cal
V}/{\cal V}_{max}=(z/z_{max})^3$. We exclude galaxies within $6^\circ$ of the
center of the Virgo cluster or $3^\circ$ of the Fornax cluster from the
$cz<1000$ km s$^{-1}$ sample. This eliminates the worst distance estimates,
although peculiar velocities clearly must affect the rest of the sample as
well. We have tested how adjustments for peculiar velocity might alter our
results using the $POTENT$ model of Dekel, Bertschinger, and Faber (1990), and
find no substantive changes from the results presented below, although the
samples generally have somewhat lower values of ${\cal V}/{\cal V}_{max}$. We
use the redshift corrected for Local Group motion here to maintain consistency
with F--T and Briggs.

With $z_{max}=z_{CZ}$, the mean value of ${\cal V}/{\cal V}_{max}$ is
$0.406\pm0.016$.  Values below 0.5 imply that galaxies were detected
preferentially in the nearer portion of the survey volume, suggesting that the
sample is not fully sensitive to sources out to $z_{CZ}$.

A low value of ${\cal V}/{\cal V}_{max}$ can alternatively be caused by an
actual clustering of galaxies nearby us. However, this tends to be
counter-balanced by the effects of morphological segregation and gas
depletion, which would favor HI detections in lower-density environments. In
any case, the F--T sample is mostly confined to within the local supercluster,
and it is not clear that there is an overall radial gradient within the
sampled region. Moreover, these effects do not explain the dependence of
${\cal V}/{\cal V}_{max}$ on HI mass.

\input tbl1.tex

In Table 1 we list the results for various galaxy samples and divide each
sample into three mass ranges.\footnote{While the upper and lower mass ranges
are unbounded on one side, the range of detected masses is approximately one
decade in both cases.} Sample 1 shows that the F--T sample in the
``completeness zone'' exhibits worse and worse completeness for lower mass
galaxies.

Using Briggs' functional form (eqn.~1) for the completeness limit, we could
increase the minimum mass for full-volume sensitivity from $10^{8.45}$ to
$10^{9.15} M_\odot$ to make ${\cal V}/{\cal V}_{max}$ close to 0.5 in all mass
ranges (sample 2 in Table 1). This revised limit would set the completeness
zone limit to $cz<330$ km s$^{-1}$ for $M_{HI}<10^8$ so that distance
uncertainty and confusion with Galactic HI would present significant problem
for an even larger portion of the sample. In addition, within such small
redshifts ${\cal V}/{\cal V}_{max}$ is probably biased upward, since the
lowest redshift sources may be lost in Galactic emission and sources detected
at redshifts below $cz=100$ km s$^{-1}$ were set to that value for the purpose
of estimating their distances. These limitations make the F--T sample highly
problematic for trying to understand properties of galaxies with HI masses
$<10^8 M_\odot$.

\section{Angular Size Limitations of Optical Samples}

Another approach to understanding the completeness of the F--T sample is to
examine the source selection criteria. F--T used a minimum angular size as
their primary selection criterion---examining spirals with diameters $a>3'$ and
Sd--Im galaxies with $a>2'$ as determined in the UGC (Nilson 1973). Because
other sources of angular diameters were also used for parts of the sample, some
smaller galaxies were also observed. To place all of the angular sizes on a
common system, we use the formulas from F--T to convert to the UGC scale.

\begin{figure}[tbh]
\plotone{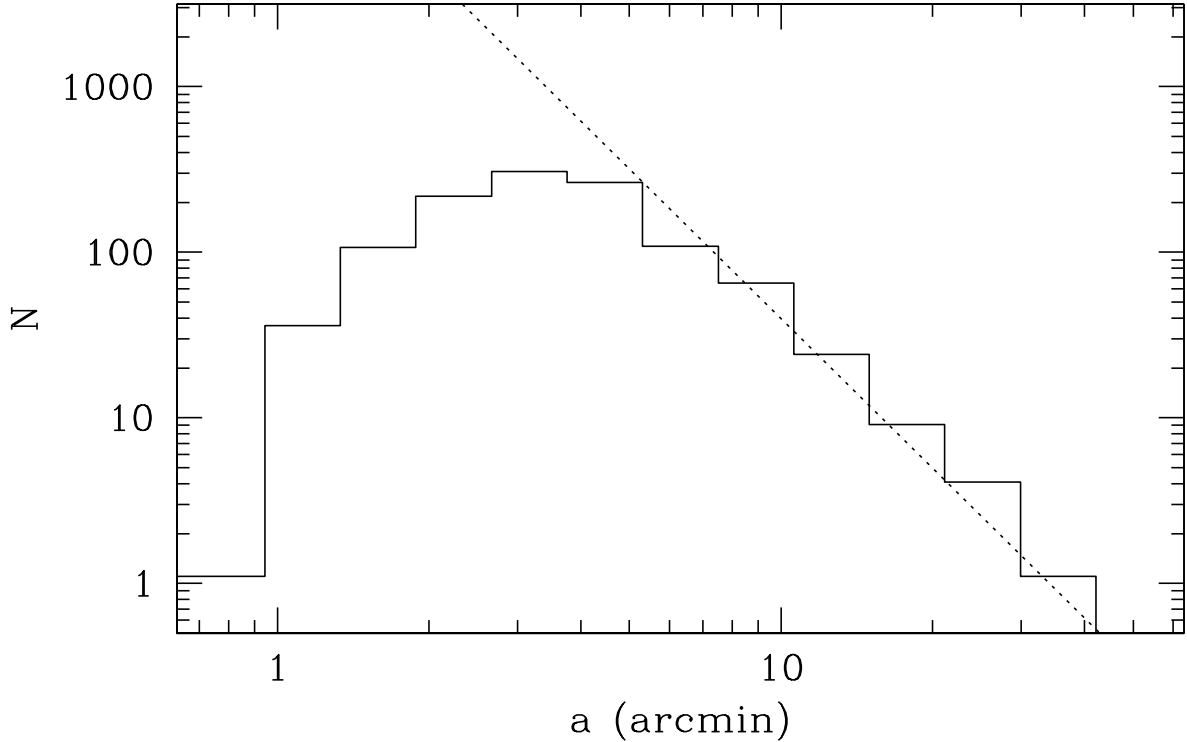}
\caption{
Histogram of F--T galaxy angular diameters. The dotted line shows the expected
slope for a complete sample distributed uniformly throughout space.
}
\label{aftu}
\end{figure}

Based on the expected increase of counts with angular diameter as $N\propto
a^{-3}$, the full sample of observed galaxies (whether or not they were
detected in HI) begins to be incomplete for angular sizes $a<4'$
(Fig.~\ref{aftu}). At $a=2'$ there are $\sim15\times$ too few galaxies relative
to the larger sources. This incompleteness at small angular sizes is partly
intentional, since F--T excluded small angular diameter galaxies that they
expected would be distant. Unfortunately, this also introduces a degree of
subjectiveness to inclusion in the sample.

Even among the F--T galaxies in Briggs' ``completeness zone,'' many of the
galaxies have angular sizes so small that they would not have remained in the
sample if they were at their maximum distance within the zone. The angular size
a source would have at $z_{CZ}$ is $a_{CZ}\equiv a\times z/z_{CZ}$. Of the 41
sources with $M_{HI}<10^8M_\odot$, none has $a_{CZ}>4'$; only one is $>3'$, the
stated completeness limit of F--T; and only 8 have $a_{CZ}>2'$. Even among the
171 intermediate mass sources, with $10^8<M_{HI}<10^9M_\odot$,  only 7\%, 25\%,
and 60\% galaxies have $a_{CZ}>4'$, $3'$, and $2'$ respectively. Only the high
mass sources are large enough that high fractions pass the $a_{CZ}$
requirement---74\%, 90\%, and 98\% for $a_{CZ}>4'$, $3'$ and $2'$. Clearly this
will tend to push ${\cal V}/{\cal V}_{max}$ to lower values since some galaxies
are included only in the near portion of the search volume.

To make a more uniform selection we can restrict the F--T sample to $a_{CZ}>3'$
(sample 3), in the intermediate and high mass ranges ${\cal V}/{\cal V}_{max}$
is below 0.5 only marginally (1.3 and 1.0 $\sigma$), but the low mass range
cannot be tested since it has only one galaxy. Restricted to $a_{CZ}>2'$
(sample 4) the F--T sample does not pass the ${\cal V}/{\cal V}_{max}$ test in
either the low or intermediate mass ranges.

One way of addressing the omission of small galaxies is to look to samples of
small galaxies. In particular the ``dwarf and LSB'' (D+LSB) sample of galaxies
from Schneider et al. (1990, 1992) contains HI measurements for late-type,
dwarf, and irregular galaxies down to a $1'$ diameter. Since this sample is
drawn from the UGC, it covers only the northern sky ($\delta>-2.5^\circ$), but
when supplemented with HI measurements from the literature (Huchtmeier \&
Richter 1989) the HI detections are more than 85\% complete. 

The ${\cal V}/{\cal V}_{max}$ test results for the D+LSB sample within Briggs'
``completeness zone'' are given in Table 1 (sample 5). The low-mass ranges
still do not pass the test, but they fare considerably better than the F--T
sample. By restricting the galaxies to $a_{CZ}>1'$, which eliminates galaxies
that only meet the UGC size criterion because they are very nearby (sample 6),
the test is passed to within $2\sigma$ in all mass ranges. This also shows
that large scale structure is not causing low ${\cal V}/{\cal V}_{max}$ test
results for the F--T sample.

We can combine the F--T and D+LSB samples in the hope of forming a complete
sample of all types of galaxies as Briggs (1997b) did. In the northern sky we
find 248 F--T galaxies and 47 additional D+LSB galaxies that satisfy the
``completeness zone'' criteria. This expanded sample fares only marginally
better in the ${\cal V}/{\cal V}_{max}$ tests, yielding $0.407\pm0.016$ for the
full sample, and in the separate mass ranges (Table 1, sample 7). Restricting
$a_{CZ}$ does not generate samples that pass the ${\cal V}/{\cal V}_{max}$ test
either.\footnote{Restricting the samples to high Galactic latitudes
($|b|>30^\circ$) made no appreciable difference to the results presented in
Table 1.} 

We conclude that the F--T sources with high HI masses represent a relatively
complete sample, but the sources with low HI masses are strongly biased to low
redshifts. The problem with low mass galaxies may be caused in part by the
F--T angular size criterion. However, even when the F--T sample is (1)
restricted to minimum physical diameters to make the galaxies relatively
uniform within the ``completeness zone,'' or (2) supplemented by galaxies from
other surveys, the galaxies with HI masses $<10^9 M_\odot$ still fail the
${\cal V}/{\cal V}_{max}$ test. This demonstrates that the F--T sample does
not provide a good basis for forming a representative cross section of galaxy
types. 

\section{High Mass LSB Galaxies}

While galaxies with high HI masses ($M_{HI}>10^9 M_\odot$) in the F--T sample
pass the ${\cal V}/{\cal V}_{max}$ test, this is really only an internal check
on the self-consistency of the database.  To examine the broader question of
how representative the F--T sample is, Briggs (1997a) asked whether surveys of
LSB galaxies had found galaxies within the ``completeness zone.'' However,
since these other surveys were also based on visual examination of
photographic plates, they do not provide a genuinely independent check of the
F--T sample. Moreover, since the local density of galaxies is higher than
average, classes of galaxies that avoid high density may not be present.

The question we consider here is whether there are massive HI sources in
deeper surveys that would have been excluded from the F--T sample because
their isophotal diameters at the PSS-I surface brightness would be less than
3--4$'$ at the 1000 km s$^{-1}$ redshift limit for high mass galaxies in the
``completeness zone.'' This is difficult to quantify precisely since diameters
estimated from the PSS-I are somewhat variable in their effective depth, but
we will adopt the mean isophotal level found by Cornell et al.~(1987) of
$\mu_{PSS-I}\equiv 25.36$ mag arcsec$^{-2}$ at $B$ for UGC diameters.

The F--T subset of high-mass galaxies are almost all physically large at
$\mu_{PSS-I}$. Figure 2 shows the distribution of $a_{CZ}$ for this (solid-line
histogram) and other samples of galaxies. Since all of the high mass galaxies
are detectable to the 1000 km s$^{-1}$ redshift limit, $a_{CZ}=1'$ corresponds
to 3.88 kpc. Thus 90\% of the F--T high mass galaxies have sizes larger than
11.6 kpc. Note that we restrict the following analyses to galaxies at high
Galactic latitudes where interstellar extinction should not much affect the
galaxies' optical sizes or number counts. \footnote{Briggs (1997ab) specifies
that his samples are restricted to high latitudes, but the numbers of galaxies
he quotes in various subsamples indicate he was using the full sky coverage
of F--T.}

\begin{figure}[p]
\epsscale{0.7}
\plotone{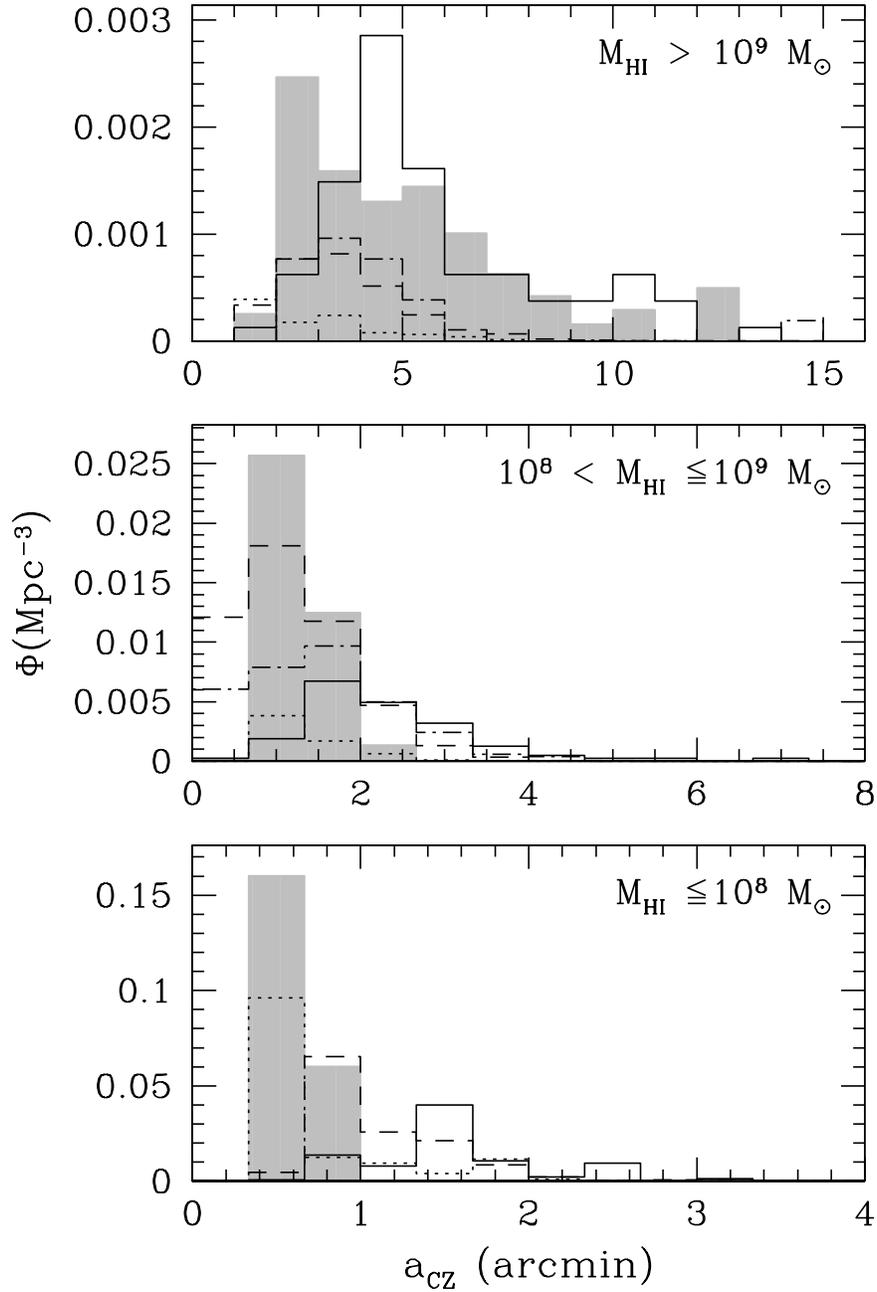}
\caption{
Number density of galaxies as a function of their angular diameter as
determined at the ``completeness zone'' distance limit. The size distribution
of the F--T sample is shown by a solid-line histogram; the UGC Dwarf+LSB
sample is shown by a dashed line, and restricted to the ``completeness zone''
by a dot-dash line; the PSS-II LSB sample is shown by a dotted line. The size
distribution of the HI-selected ``Slice'' sample is shown by a solid gray
histogram. The densities of the F--T sample and the D+LSB sample in the
low-mass range are divided by two to account for the local overdensity as
explained in the text. 
}
\label{massacz}
\end{figure}

We give the estimated number density of each size of galaxy in the figure
based on the areal coverage of the sample ($\sim$5.1 sr for the high-latitude
portion of the F--T sample). We also need to account for the local overdensity
of galaxies when making comparisons to other samples. Briggs (1997b) estimates
the region inside $cz<1000$ km s$^{-1}$ has a density about a factor of 2
above average. This matches our results for the D+LSB sample, which has a
density 2.1 times higher in the nearby portion. Densities in Fig.~2 that are
based on the F--T sample and other samples restricted to the local region are
divided by 2.

The ``HI-Slice'' sample of Spitzak \& Schneider (1998) was found by
systematically observing 55 sq deg of the sky at 21 cm from Arecibo, and is
unbiased by optical sizes. This survey contains 62 sources with $M_{HI}>10^9
M_\odot$. We have determined the angular sizes at $\mu_{PSS-I}$ from the
original $B$-band photometric profile fits of Spitzak \& Schneider. Compared to
the F--T sample, these galaxies have lower percentages of optically large
galaxies, and higher percentages of small galaxies, although there appear to be
very few galaxies smaller than 7.8 kpc in either sample.

Using the sensitivity limit estimates from Schneider, Spitzak, \& Rosenberg
(1998), we can determine the volume within which each of the HI-Slice galaxies
was detectable in order to estimate its number density in space. The results
are shown in Fig.~2 by the solid-gray histogram. We estimate that 75\% of the
HI among these massive galaxies is associated with galaxies larger than
$a_{CZ}>4'$. Assuming the F--T sample is complete for these largest galaxies,
the smaller fraction of galaxies it finds at smaller sizes implies it is
missing about 23\% of high-mass galaxies and 12\% of the total HI due to the
angular size limitations.

Several differences between the samples may reflect environmental influences.
The distribution of F--T galaxy sizes in Fig.~2 suggests that the population
has been shifted to systematically larger sizes than the HI-Slice galaxies. In
addition, the HI-selected sample has $1.8\times$ less $B$-band luminosity
relative to the HI mass on average, based on optical data from Spitzak \&
Schneider and the RC3 (de Vaucouleurs et al. 1991), and 40\% of the HI-selected
galaxies have $M_{HI}/L_{B}>1$ (in solar units) compared to 11\% of the F--T
galaxies. Since the F--T sample is located in a region of high galaxy
density, star formation induced by galaxy interactions may explain these
differences.

By contrast, {\it optically}-selected samples, even of LSB galaxies, rarely
identify high HI mass objects that would not have been identified in Briggs'
``completeness zone.'' If we impose no HI-mass/redshift restrictions, the
D+LSB sample (\S 3) contains 586 high-HI-mass sources with $|b|>30^\circ$, and
it has even higher fractions of large-$a_{CZ}$ galaxies than the F--T sample.
This suggests that almost all UGC galaxies with large HI masses have large
physical dimensions at $\mu_{PSS-I}$.

However, the raw distribution of $a_{CZ}$ in an angular-diameter limited
sample is not a good indicator of the population size distribution since
galaxies with smaller physical diameters at $\mu_{PSS-I}$ remain larger than
$1'$ (the UGC limit) out to smaller distances. Dividing the counts by the
volume in which each galaxy remains larger than $1'$, the distribution has a
similar shape to the HI-selected sample (dashed-line histogram in Figure 2).
These galaxies also appear to have distinctly different properties from
galaxies in the F--T sample---their $B$-band luminosities relative to their HI
masses are similar to the HI-selected sample of sources. 

The distribution of sizes of the subset of D+LSB galaxies within $cz<1000$ km
s$^{-1}$ is shown by a dot-dash line. The counts have been divided by 2 (like
the F--T sample) and demonstrate that the density correction used earlier is
reasonable. The nearby D+LSB galaxies show a slight shift toward larger sizes
than the full sample. This indicates that the distribution of sizes in a
sample of late-type galaxies alone is little-affected by the local density
enhancement. By contrast, the F--T and HI-selected samples contain a wide
range of morphological types. Since earlier-type galaxies tend to be larger at
the same isophotal level, the shift in the size distribution to larger
$a_{CZ}$ within the F--T sample may reflect the effects of morphological
segregation, with a larger proportion of earlier types than an average sample.

Within the northern portion of the ``completeness zone'' the D+LSB sample adds
only 1 high-mass galaxy to the F--T sample, so the size distribution of the
combined sample is basically unchanged. The lack of additional
small angular diameter galaxies supports the idea that the local population is
different from samples drawn from a wider variety of environments.

The PSS-II survey of Eder \& Schombert (1999) is a deeper probe for small
diameter, very late-type galaxies. These objects were selected from 50 PSS-II
plates ($\sim$0.7 sr) to be larger than 20$''$ and to have dwarf-like
morphologies. We have determined the sizes on the PSS-I for each of these
objects; our size estimates are consistent for galaxies in common with the UGC.
The size distribution turns over below diameters of 0.4$'$ at $\mu_{PSS-I}$,
which we take to be the effective completeness limit.

While aimed primarily at identifying dwarf galaxies, the PSS-II sample includes
135 galaxies with $M_{HI}>10^9M_\odot$. Their size distribution is shown by a
dotted line in Fig.~2. Only 7 of these objects are smaller than $a_{CZ}<2'$,
but based on their angular size distance limits, such small sources comprise
about half of the population of high mass LSB objects with very late type
morphologies.

Extremely high HI mass LSB systems like Malin 1 (Bothun et al.~1987) might
have been missed in the nearby volume of space because their disks are so
faint that even the central extrapolated surface brightness of the disk is
fainter than $\mu_{PSS-I}$. However, the bulge component of Malin 1 would
reach $\mu_{PSS-I}$ at about $3.75'$ at $cz=1000$ km s$^{-1}$. We can
speculate that a system like Malin 1 might have been identified as an E or S0
in the UGC, and therefore omitted from consideration for the F--T sample. It
is worth noting that a number of early-type galaxies have been found with
extended distributions of HI (Van Driel \& Van Woerden 1991; DuPrie \&
Schneider 1996), and perhaps these are the more appropriate comparison to
Malin 1. For other giant disk systems that have been compared to Malin 1, like
F568-6 (Bothun et al.~1990) and 1226+0105 (Sprayberry et al.~1993), the disk
surface brightness and size would make these objects easily exceed
$a_{CZ}>3'$, and should thus have been included in the F--T sample if they
were in the ``completeness zone.''

In summary, high-HI-mass galaxies are relatively well-sampled by F--T, but
they do miss an interesting fraction of galaxies that have small sizes at the
PSS-I isophotal limit. Based on intrinsic differences in the optical-to-21 cm
emission from the nearby F--T sample versus more-distant samples, we suggest
that the high density of galaxies in the local environment may cause
differences in the local population.

\section{Low and Intermediate Mass Galaxies}

Galaxies with less than $10^9 M_\odot$ of HI are highly incomplete in the F--T
sample. Among the galaxies identified by F--T in this mass range, about half
have $a_{CZ}<2'$, and only $7\%$ are larger than $a_{CZ}>4'$, so F--T's
adopted angular size constraints give rise to a fundamental limitation to the
sample's completeness. We illustrate here the degree of the incompleteness and
attempt to extrapolate to the population of missing objects.

The distribution of sizes in the F--T sample among lower mass galaxies is shown
in the bottom two panels of Fig.~2, divided into intermediate
($10^8-10^9M_\odot$) and low ($<10^7M_\odot$) HI masses. Since $z_{CZ}$
declines for galaxies with HI masses $<10^{8.45}M_\odot$ (eqn.~1), the density
is estimated from the corresponding volume. At $10^8 M_\odot$, the distance
limit and $a_{CZ}$ are 65\% of their value for high-mass galaxies, and at $10^7
M_\odot$ they are 25\% as big.

The HI-Slice sample of Spitzak \& Schneider (1998) contains 3 low and 10
intermediate mass galaxies. Despite the small-number statistics, this sample
clearly demonstrates that very small sizes are the norm among low-HI-mass
galaxies as shown in Fig.~2. {\it All} of the sources are smaller than
$a_{CZ}=2.2'$, and 5 of 6 sources with $M_{HI}<2.5\times10^8$ are smaller than
$a_{CZ}<1'$. One relatively high mass source ($M_{HI}=7\times10^8M_\odot$) has
$a_{CZ}=0.13'$ and is nearly invisible on the PSS-I.

The size-distribution of the HI-Slice galaxies is clearly different from the
F--T sample, exhibiting a strong peak toward the smallest diameters. The
lowest-mass source in the HI-Slice sample (\#75 in Spitzak \& Schneider) was
not detected optically because of a foreground star, but it is certainly
very small. Given its low potential detection volume, it would increase the
estimated density of the smallest-size low-mass galaxies by more than a factor
of six. Since its contribution is not included in the histogram, the density of
very small low-mass galaxies may be substantially larger than shown.

The D+LSB sample contains 482 galaxies with $M_{HI} < 10^9 M_\odot$. 22 of
these galaxies have $a_{CZ}<1'$, ranging down to $0.44'$ and physical diameters
as small as 0.7 kpc at the UGC isophote. All 22 of these small galaxies fall
within the F--T ``completeness zone'' but such galaxies would be overlooked
even in the combined F--T and D+LSB samples if they were beyond the nearest
portion of the zone.

Most of the D+LSB galaxies are larger than $2'$, but after adjusting for each
source's maximum detectable distance according to its angular size, we find the
density distribution shown by the dashed-line histogram in Fig.~2. In the
intermediate mass range this distribution is quite similar to the HI-Slice
sample for galaxies with small diameters.

The portion of the D+LSB sample restricted to the ``completeness zone'' (24\%
of 394 galaxies) is again shown with a dot-dash histogram in Fig.~2. The size
distribution of these galaxies begins to resemble that of the F--T sample even
though small-$a_{CZ}$ galaxies would be easier to detect nearby. This suggests
again that the nearby volume of space is atypical. 

All three of the low-mass HI-Slice galaxies and 92\% of the 88 low-mass D+LSB
galaxies have redshifts below $cz=1000$ km s$^{-1}$, although most are outside
the ``completeness zone'' distance limit at these masses. We have divided the
densities of the low-mass D+LSB galaxies by a factor of 2 as we did for the
F--T sample since it mostly probes the same volume of space. The HI-Slice
sample density estimates are already adjusted for the local large-scale
structure in the direction of that survey (Schneider et al. 1998).

We estimate the space density of small, LSB galaxies based on 23 low-mass and
77 intermediate-mass galaxies that are larger than $0.4'$ in the PSS-II sample
of Eder \& Schombert (1999). Most of the low-mass galaxies are within $cz<1000$
km s$^{-1}$, so the densities should perhaps be divided by 2; however,
the sample is partially incomplete for the smallest sizes, so the densities may
be underestimated. The densities (dotted histogram in Fig.~2) assume that
sources were detectable out to the distance where their angular diameter would
reach $0.4'$.

The lowest-mass galaxy in the Eder \& Schombert LSB sample (D634-3) was not
included in the density estimates. For this galaxy $V_0=181$ km s$^{-1}$, so
its distance and mass are quite uncertain. Taken at face value this source
would imply a density of 0.7 Mpc$^{-3}$ of very low-mass objects, comparable to
the large density implied by the lowest-mass source in the HI-Slice sample. 

In summary, optically-selected samples of galaxies only begin to indicate the
prevalance of small-diameter galaxies as measured at the limiting isophotal
depth of the PSS-I. Photographic surveys of galaxies with late-type
morphologies can recover the density of objects with intermediate HI masses if
the selection criteria are well-understood, but low mass galaxies present a
much bigger challenge. Detections of two very small, low mass galaxies in an HI
survey and an LSB survey imply that there may be a very large population of
sources with $M_{HI}<10^7M_\odot$, but the statistical uncertainties are too
great to draw firm conclusions on this point.

\section{Discussion}

Optically selected samples favor optically bright galaxies. This truism holds
even for diameter-limited galaxy surveys because LSB galaxies appear small at
the surface-brightness limit of the optical images (Disney 1976). The HI-Slice
survey (Spitzak \& Schneider 1998), which is unbiased by galaxy diameter,
indicates that the optically smallest galaxies are the most common. Current
searches on deep photographic plates for small angular diameter sources (Eder
\& Schombert 1999) are also uncovering indications of this population. Such
LSB and 21 cm surveys are successfully probing sources with HI masses down to
$\sim10^8M_\odot$, but for lower mass objects HI flux and angular size
limitations of existing surveys allow detections of these sources to only a
few Mpc.

Because of the small distances at which low mass and LSB sources are
accessible, we need to consider the possible impact of the local environment
on them. Although the large scale distribution of LSB galaxies appears similar
to that of other galaxies (Mo, McGaugh and Bothun 1994), on scales of less
than 2 Mpc their numbers drop off sharply. The most likely explanations are
that either LSB disks are fragile and easily disturbed by other galaxies, or
tidal interactions induce star formation that converts LSB galaxies into
normal Hubble type objects. Regardless of the underlying cause why LSB
galaxies avoid high density regions, this fact produces an expectation that
the local region of space will be deficient in the number of LSB disk galaxies
due to the large number of high mass spirals and the proximity to the very
dense Virgo Cluster.

Other influences of the local environment may also play a role in the
distribution of galaxy types. In high density regions, galaxies are often gas
deficient for their morphological type because of stripping or evaporation,
and morphological segregation favors earlier-type, less gas rich galaxies.
Both of these effects would tend to {\it lower} the percentage of gas-rich
systems nearby.

Briggs (1997a) asked ``Where are the nearby gas-rich low surface brightness
galaxies?'' The answer depends on the mass range of objects being studied.
There appears to be a local deficit of high-mass LSB systems, which is
probably an environmental effect. The story for low-mass systems is less
settled, because of the difficulty in detecting them to any significant
distance, but it is clear there is a much larger population of
small-optical-diameter galaxies than optical surveys have previously revealed.
And finally, since these low-mass objects have not yet been detected beyond
the local high-density environment, it is possible that they are even more
profuse than they appear locally.

\begin{acknowledgments}

\end{acknowledgments}

\end{document}

%% file: tbl1.tex
\clearpage

\begingroup
\oddsidemargin -.5in
\evensidemargin -.5in
 
\begin{deluxetable}{@{}l@{\quad}r@{ }r@{\quad }r@{ }r@{\quad }r@{ }r@{\quad }r@{ }r}
\renewcommand{\arraystretch}{0.82}
\tablecolumns{9}
\tablewidth7.3in
\footnotesize
\tablecaption{${\cal V}/{\cal V}_{max}$ Tests. \label{tbl-1}}
\tablehead{
&\multicolumn{2}{c}{\hbox to 1.1in{\hss All\quad\hss}}
&\multicolumn{2}{c}{\hbox to 1.1in{\hss $M_{HI}\leq10^8 M_\odot$\quad\hss}}
&\multicolumn{2}{c}{\hbox to 1.1in{\hss $10^8-10^9 M_\odot$\quad\hss}}
&\multicolumn{2}{c}{\hbox to 1.1in{\hss $M_{HI}>10^9 M_\odot$\quad\hss}}\\
Sample&
\colhead{$N$}&\colhead{${\cal V}/{\cal V}_{max}$} &
\colhead{$N$}&\colhead{${\cal V}/{\cal V}_{max}$} &
\colhead{$N$}&\colhead{${\cal V}/{\cal V}_{max}$} &
\colhead{$N$}&\colhead{${\cal V}/{\cal V}_{max}$}
} 
\startdata
1. F--T in ``completeness zone''                         &320&$0.406\pm0.016$& 41&$0.299\pm0.046$&171&$0.392\pm0.022$&108&$0.468\pm0.028$\nl
2. $\ldots$full sensitivity to $M_{HI}>10^{9.15} M_\odot$&190&$0.485\pm0.021$& 15&$0.510\pm0.075$& 69&$0.481\pm0.035$&106&$0.484\pm0.028$\nl
3. $\ldots$restricted to $a_{CZ}>3'$                     &141&$0.460\pm0.024$&  1&$0.028\pm0.287$& 43&$0.443\pm0.044$& 97&$0.472\pm0.029$\nl
4. $\ldots$restricted to $a_{CZ}>2'$                     &217&$0.429\pm0.020$&  8&$0.208\pm0.102$&103&$0.404\pm0.028$&106&$0.469\pm0.028$\nl
5. D+LSB in ``completeness zone''                        &189&$0.437\pm0.021$& 52&$0.425\pm0.040$&116&$0.425\pm0.027$& 21&$0.539\pm0.063$\nl
6. $\ldots$restricted to $a_{CZ}>1'$                     &161&$0.462\pm0.023$& 36&$0.459\pm0.048$&104&$0.448\pm0.028$& 21&$0.539\pm0.063$\nl
7. F--T ($\delta>-2.5^\circ$) and D+LSB combined         &295&$0.407\pm0.016$& 54&$0.385\pm0.039$&162&$0.416\pm0.023$& 79&$0.476\pm0.032$\nl
8. $\ldots$restricted to $a_{CZ}>2'$                     &172&$0.432\pm0.022$&  6&$0.161\pm0.118$& 88&$0.410\pm0.031$& 78&$0.478\pm0.033$\nl
9. $\ldots$restricted to $a_{CZ}>1'$                     &270&$0.435\pm0.018$& 40&$0.382\pm0.046$&151&$0.427\pm0.023$& 79&$0.476\pm0.032$\nl
\enddata
\end{deluxetable}

\clearpage
\endgroup